\def\cm#1{}

\newcommand{\deltabar}{\,\,{\bar{}\hspace{1pt}\! \!\delta }}

\documentstyle[12pt,appb]{article}

\begin{document}
\title{\Large\bf Novel Geometric Gauge Invariance
\\of Autoparallels}
\author{H. Kleinert%
\thanks{
kleinert@physik.fu-berlin.de,
http://www.physik.fu-berlin.de/\~{}kleinert }
\, and A. Pelster%
\thanks{
pelster@physik.fu-berlin.de,
http://www.physik.fu-berlin.de/\~{}pelster }
  \\ [3mm]
{\em Institut f\"ur Theoretische Physik,}\\
{\em Freie Universit\"at Berlin, Arnimallee 14,
D-14195 Berlin, Germany}\\ [5mm] }
\maketitle
\begin{abstract}
We draw attention to a novel type of geometric gauge invariance
relating the autoparallel
equations of motion in different Riemann-Cartan spacetimes
with each other.
The novelty lies in the fact that
the equations of motion are invariant even though
the
actions are not.
As an application
we use this
gauge transformation to map the action of a spinless point particle
in a Riemann-Cartan spacetime with a gradient torsion to
a purely Riemann spacetime, in which
the initial torsion
appears as a nongeometric external field.
By extremizing the transformed action in the usual way, we obtain
the same autoparallel equations of motion
as
those derived in the initial
spacetime with torsion via a recently-discovered variational principle.
\end{abstract}

\section{Introduction}

Einstein's theory of general relativity
predicts correctly all post-Newtonian
experiments in our solar system as well
as some effects of strong
gravitational fields
observed in
binary
systems of
neutron stars \cite{Taylor}.
The theory has, however, two unsatisfactory properties.
One is the somewhat academic
fact that gravity cannot be quantized
in a renormalizable way
\cite{Goroff}, but only as
an effective theory which is unable to
predict short-distance gravitational phenomena
at a scale much shorter than the Planck length
$l_P \approx 10^{-32} $cm \cite{Weinberg1}.
 Since this length scale is extremely small,
the problem is not very serious for present-day physics.
The other unsatisfactory property
which has received much attention
is of an esthetical nature.
Since all elementary-particle forces known so far
are mediated by local gauge fields, one would like to
describe also the geometric theory
of gravity as a local
Poincar\'e gauge symmetry.
This would naturally
introduce
torsion into the geometry \cite{Utiyama,Kibble,Hehl1,Kleinert1}, thus
extending Riemann spacetime of the Einstein theory to
Riemann-Cartan spacetime.
Therein
parallelograms
exhibit
both an angular
and a closure failure
due to curvature and torsion, respectively.
So far, such spacetimes
have only found applications
in the theory of
plastic flow
and material fatigue
\cite{Kleinert1,Bilby,Kroener1,Kroener2}, where
curvature and torsion
are produced by
disclinations and dislocations.\\

The last two decades have seen
a detailed
elaboration of the gauge-theoretic
formulations
of gravity in
spacetimes with torsion, most notably the
Einstein-Cartan theory \cite{Hehl1,Hehl2}. The latter has the
appealing
feature
that Einstein's equation
stating the proportionality between curvature and
energy momentum tensor of matter is extended by a corresponding one
involving torsion and spin density. In this theory
spinless particles do not create
torsion,  and
since  usually only particles
which are sources
of a field can also be influenced by this field,
it
has generally been
believed
that trajectories of spinless
particles
are
not
affected by torsion in spacetime, i.e.,
that
spinless point
particles move along
{\it geodesics},
the shortest curves in the spacetime.
With this bias
the relativity community
happily embraced Hehl's derivation \cite{Hehl3}
of such trajectories from hitherto accepted gravitational field
equations as has been recently reviewed in \cite{Sabbata}.\\

This belief and its field theoretic derivation
have recently been challenged
by one of us
\cite{Kleinert2,Kleinert3,Kleinert4,Kleinert7,Kleinert5,Kleinert6,Shabanov,moregen}.
It
was pointed out that the invariance under general coordinate transformations
of general
relativity
used by Einstein to find
the laws of nature in a curved geometry may be replaced by
a more efficient {\it nonholonomic mapping principle},
which is moreover predicitve
for spacetimes with torsion.
This new
principle was originally discovered
for the purpose
of
transforming
nonrelativistic path integrals correctly from flat space
to spaces with torsion \cite{Kleinert4}, where it played the role
of
a
{\em quantum equivalence
principle\/}.
Evidence for its correctness was derived from
its essential role
in solving the path integral
of the hydrogen atom, where  the nonholonomic
Kustaanheimo-Stiefel transformation
was used
\cite{Kleinert4}.\\

Applying the nonholonomic mapping principle to the variational procedure
has the important consequence that
the Euler-Lagrange equations of spinless point particles receive
an additional force which depends on the
torsion tensor $S_{\mu\nu}^{\,\,\,\,\,\,\lambda} ( q )$
\cite{Kleinert6}. If $\tau$ denotes an arbitrary parameter of
the trajectory $q^{\lambda} ( \tau )$, the modified Euler-Lagrange equations
read
\begin{equation}
\label{EL3}
\frac{d}{d \tau } \, \frac{\partial L}{\partial
\dot{q}^{\lambda} ( \tau  )} -
\frac{\partial L}{\partial q^{\lambda} ( \tau  )}
= - 2  S_{\lambda\mu}^{\,\,\,\,\,\,\nu} \left( q ( \tau  ) \right)
\dot{q}^{\mu} ( \tau  ) \frac{\partial
L}{\partial \dot{q}^{\nu} ( \tau  )}
\, .
\end{equation}
The modification on the right-hand side has its origin
in the closure failure of parallelograms in spacetimes with torsion,
which can be accounted for by a noncommutativity of
nonholonomic variations
$\deltabar$ with the
parameter derivative $d_{\tau} = d / d \tau$
of the trajectory $q^{\lambda} ( \tau )$ \cite{Kleinert6}:
\begin{equation}
\label{VAR7}
\deltabar
d_{\tau } q^{\lambda} ( \tau  ) - d_{\tau } \deltabar
 q^{\lambda} ( \tau  )
= 2 S_{\mu\nu}^{\,\,\,\,\,\,\lambda} ( q ( \tau  ) ) \dot{q}^{\mu}
( \tau ) \deltabar q^{\nu} ( \tau  ) \, .
\end{equation}
For the free-particle Lagrangian \cite{Weinberg2}
\begin{equation}
L ( q^{\lambda}, \dot{q}^{\lambda} ) = - Mc \,
 \sqrt{g_{\lambda\mu} ( q ) \dot{q}^{\lambda}
\dot{q}^{\mu} }.
 \label{AC3}\end{equation}
with the proper time
\begin{equation}
\label{PT2}
d s ( \tau  ) =
 \sqrt{g_{\lambda\mu} ( q ( \tau  ) ) d q^{\lambda} ( \tau  )
d q^{\mu} ( \tau  )} \, ,
\end{equation}
the modified Euler-Lagrange equation (\ref{EL3})
becomes
explicitly
\begin{eqnarray}
\ddot{q}^{\lambda} ( s  ) +
g^{\lambda\kappa} ( q ( s ) )
\left[ \partial_{\mu} g_{\nu\kappa} ( q ( s  ) ) - \frac{1}{2}
\partial_{\kappa} g_{\mu\nu} ( q ( s  ) ) \right]
\dot{q}^{\mu} ( s ) \dot{q}^{\nu} ( s ) \nonumber \\
= - 2 S_{\mu\nu}^{\,\,\,\,\,\,\lambda} ( q ( s  ) )\dot{q}^{\mu}
( \tau  ) \dot{q}^{\nu} ( s  ) \, . \hspace*{3cm} \label{ZW}
\end{eqnarray}
Here we can use the decomposition of the affine connection
$\Gamma_{\mu\nu}^{\,\,\,\,\,\,\lambda} ( q )$
in a Riemann-Cartan spacetime \cite{Schouten}
\begin{equation}
\label{AF1}
\Gamma_{\mu\nu}^{\,\,\,\,\,\,\lambda} ( q ) =
\overline{\Gamma}_{\mu\nu}^{\,\,\,\,\,\,\lambda} ( q )
+ K_{\mu\nu}^{\,\,\,\,\,\,\lambda} ( q ) \, ,
\end{equation}
where the first term is the Christoffel connection
\begin{equation}
\label{CH1}
\overline{\Gamma}_{\mu\nu}^{\,\,\,\,\,\,\lambda} ( q ) =
\frac{1}{2} g^{\lambda\kappa} ( q ) \Big[ \partial_{\mu} g_{\nu\kappa} ( q )
+ \partial_{\nu} g_{\kappa\mu} ( q ) - \partial_{\kappa} g_{\mu\nu} ( q )
\Big]
\end{equation}
depending only on the metric $g_{\mu\nu} ( q )$,
while the second one is the contortion
tensor
\begin{equation}
\label{KO1}
K_{\mu\nu}^{\,\,\,\,\,\,\lambda} ( q ) =
S_{\mu\nu}^{\,\,\,\,\,\,\lambda} ( q ) -
S_{\nu\,\,\,\,\mu}^{\,\,\,\lambda} ( q ) +
S^{\lambda}_{\,\,\,\,\mu\nu} ( q ) \, ,
\end{equation}
representing a combination of the
torsion tensor $S_{\mu\nu}^{\,\,\,\,\,\,\lambda} ( q )$. Then
(\ref{ZW}) reduces to the straightest lines
or
{\em autoparallels\/} in a Riemann-Cartan spacetime:
\begin{equation}
\label{AE1}
\ddot{q}^{\lambda} ( s  ) +
\Gamma_{\mu\nu}^{\,\,\,\,\,\,\lambda} ( q ( s  ) ) \dot{q}^{\mu}
( s ) \dot{q}^{\nu} ( s  ) = 0 \, .
\end{equation}
The coupling of
spinless point particles to torsion in (\ref{AE1}) has a single
geometrical interpretation. To this end we observe that
the natural covariant derivative of an arbitrary
vector field $V^ \lambda( q )$
\begin{eqnarray}
D_\mu V^ \lambda(q )=\partial _\mu V^ \lambda(q)+
\Gamma_{\mu \nu}{}^{\lambda} ( q )  V^{\nu} ( q )
\end{eqnarray}
contains the full affine connection (\ref{AF1})
rather then the Christoffel connection (\ref{CH1}). Thus
we may define
a covariant derivative of $ V^ \lambda(q (s )) $
with
respect to the proper time $s$ as
\begin{equation}
\label{CO}
\frac{D}{D s } V^{\lambda} ( q ( s  ) ) =  \frac{d}{d s }
V^{\lambda} ( q ( s  ) ) +  \Gamma_{\mu\nu}^{\,\,\,\,\,\,\lambda} ( q
( s  ) ) \dot{q}^{\mu} ( s  ) V^{\nu} ( q ( s  ) )  \,,
\end{equation}
so that the autoparallel equation (\ref{AE1})
reads simply
\cite{Schroedinger}
\begin{equation}
\label{A4}
\frac{D}{D s} \dot{q}^{\lambda} ( s  )
=0 \,.
\end{equation}
Therefore the transition from the Minkowski to the Riemann-Cartan
spacetime corresponds to the substitution of the total derivative
$d / d s$ by the covariant one $D / D s$. This means that the free
spinless point particle is minimally coupled to the gravitational field
in a Riemann-Cartan spacetime.

\section{Novel Type of Geometric Gauge Invariance}

We now turn to the essential
point of our lecture that
the new variational procedure \cite{Kleinert6}
implies a novel kind of geometric gauge
invariance of the
autoparallel trajectories although the action is {\em not} invariant.
It turns out that a change in the action can be compensated by a
corresponding change
in the closure failure at the endpoints of the paths.
This gauge invariance is based on transformations
which were introduced in 1982 by two different research groups
in a different context \cite{Obukhov,Tucker}.
Following their notation,
we transform
metric and torsion
simultaneously, the first conformally,
the second
by adding a gradient term. The Riemann-Cartan curvature tensor
remains invariant under this transformation
which merely shifts part of the geometry
from the Riemannian to the torsion part.
We shall demonstrate that although the particle
action changes under this transformation,
initial and final actions yield the same
autoparallel particle trajectory, thus making the two geometries
indistinguishable from each other by any measuring process
involving spinless point-like test particles.
In particular the  gauge invariance will allow us to relate the action
of point particles
in a specific family of Riemann-Cartan spacetimes
in which torsion arises from the gradient of a scalar field
to that  in a Riemann spacetime. In the transformed action, the scalar field
plays the role of a nongeometric external field,
and particle trajectories can  be derived
via the traditional action principle,
yielding the same autoparallels as
in the initial spacetime with gradient torsion
via the new action principle.

\subsection{Geometry Transformations}

In a Riemann-Cartan geometry \cite{Schouten,Schroedinger},
the metric $g_{\mu\nu} ( q )$ and the affine
connection $\Gamma_{\mu\nu}^{\,\,\,\,\,\,\lambda} ( q )$
are not independent of each other.
The decomposition (\ref{AF1})--(\ref{KO1}) implies that
they must satisfy
the metricity
condition
\begin{equation}
\label{MET1}
D_{\lambda} g_{\mu\nu} ( q ) = \partial_{\lambda} g_{\mu\nu} ( q ) -
\Gamma_{\lambda\mu}^{\,\,\,\,\,\,\kappa} ( q ) g_{\kappa\nu} ( q ) -
\Gamma_{\lambda\nu}^{\,\,\,\,\,\,\kappa} ( q ) g_{\mu\kappa} ( q ) = 0 \, .
\end{equation}
The metric $g_{\mu\nu} ( q )$ and the torsion tensor
$S_{\mu\nu}^{\,\,\,\,\,\,\lambda} ( q )$, however, represent independent
quantities characterizing the Riemann-Cartan geometry.\\

The fundamental gauge invariance
of a Riemann-Cartan geometry contains two ingredients \cite{Obukhov,Tucker}.
Following Weyl
\cite{Weyl}, who postulated
that no physical phenomenon
should depend on the choice of dimensional units,
we
transform
the metric conformally  to
\begin{equation}
\label{M}
\tilde{g}_{\mu\nu} ( q ) = e^{ 2 \sigma ( q )} {g}_{\mu\nu} ( q )
\end{equation}
with an arbitrary scalar function $\sigma ( q )$.
We supplement
this transformation
by a change of
the torsion tensor according to
\begin{equation}
\label{S}
\tilde{S}_{\mu\nu}{}^{\lambda} ( q ) =
{S}_{\mu\nu}{}^{\lambda} ( q ) + \frac{1}{2}
\left[ \delta_{\nu}^{\,\,\,\,\lambda} \partial_{\mu} \sigma ( q ) -
\delta_{\mu}^{\,\,\,\,\lambda} \partial_{\nu} \sigma ( q ) \right] \, .
\end{equation}
For the
Christoffel connection (\ref{CH1})
and the contortion tensor (\ref{KO1}),
these transformations imply
\begin{eqnarray}
\label{CH2}
\tilde{\overline{\Gamma}}_{\mu\nu}{}^{\lambda} ( q ) & = &
{{\overline{\Gamma}}}_{\mu\nu}{}^{\lambda}
( q ) +
\delta_{\nu}^{\,\,\lambda} \partial_{\mu} \sigma ( q ) +
\delta_{\mu}^{\,\,\lambda} \partial_{\nu} \sigma ( q )
- g_{\mu\nu} ( q ) g^{\lambda\kappa} ( q ) \partial_{\kappa}
\sigma ( q ), \hspace*{0.6cm} \\
\label{KO2}
\tilde{K}_{\mu\nu}{}^{\lambda} ( q ) & = &
{K}_{\mu\nu}{}^{\lambda} ( q )
- \delta_{\mu}^{\,\,\lambda} \partial_{\nu} \sigma ( q )
+ g_{\mu\nu} ( q ) g^{\lambda\kappa} ( q ) \partial_{\kappa}
\sigma ( q ),
\end{eqnarray}
respectively.
In the affine
connection (\ref{AF1}),
the transformations (\ref{CH2}), (\ref{KO2})
almost compensate each other, resulting
only in
an additional gradient term:
\begin{equation}
\label{AF2}
\tilde{\Gamma}_{\mu\nu}{}^{\lambda} ( q ) =
{\Gamma}_{\mu\nu}{}^{\lambda} ( q ) +
\delta_{\nu}^{\,\,\,\,\lambda} \partial_{\mu} \sigma ( q ) \, .
\end{equation}
In
the Riemann-Cartan geometry,
this leaves
the curvature tensor
\begin{equation}
R_{\mu\nu\kappa}^{\,\,\,\,\,\,\,\,\,\,\lambda} ( q )  = \partial_{\mu}
\Gamma_{\nu\kappa}^{\,\,\,\,\,\,\lambda} ( q ) - \partial_{\nu}
\Gamma_{\mu\kappa}^{\,\,\,\,\,\,\lambda} ( q ) +
\Gamma_{\nu\kappa}^{\,\,\,\,\,\,\rho} ( q )
\Gamma_{\mu\rho}^{\,\,\,\,\,\,\lambda} ( q ) -
\Gamma_{\mu\kappa}^{\,\,\,\,\,\,\rho} ( q )
\Gamma_{\nu\rho}^{\,\,\,\,\,\,\lambda} ( q )
\end{equation}
invariant:
\begin{equation}
\tilde{R}_{\mu\nu\kappa}{}^{\lambda} ( q ) =
{R}_{\mu\nu\kappa}{}^{\lambda} ( q ).
\end{equation}
The associated curvature scalar,
\begin{equation}
R ( q ) = g^{\nu\kappa} ( q )
R_{\mu\nu\kappa}^{\,\,\,\,\,\,\,\,\,\,\mu} ( q ),
\end{equation}
however, is changed by
a conformal
transformation of the inverse metric following from (\ref{M}),
so one obtains
\begin{equation}
\tilde{R} ( q ) = e^{- 2 \sigma ( q )} {R} ( q ) \, .
\end{equation}
Due to (\ref{AF2}), the geometry  transformations (\ref{M}), (\ref{S})
change covariant
derivatives of the metric by a conformal factor
\begin{equation}
\tilde{D}_{\lambda}
\tilde{g}_{\mu\nu} ( q ) = e^{2 \sigma ( q )} \,
{D}_{\lambda} g_{\mu\nu}
( q ) \, .
\end{equation}
As a consequence, covariant
derivatives with respect to
general coordinate transformations
also remain covariant under the geometry  tranformations.
For this reason,
the metricity condition
(\ref{MET1}) holds also after a geometry  transformation
\begin{equation}
\label{MET2}
\tilde{D}_{\lambda} \tilde{g}_{\mu\nu}
( q ) =  0 \, ,
\end{equation}
guaranteeing the gauge invariance of the Riemann-Cartan geometry.

\subsection{Application to Autoparallel Particle Trajectories}

We now prove our principal result
that the pair of geometry transformations
(\ref{M}) and (\ref{S}) between
different Riemann-Cartan spacetimes
leaves trajectories
of spinless point particles invariant.
To this end we start with the Lagrangian
\begin{equation}
\tilde{L} ( q^{\lambda}, \dot{q}^{\lambda} ) = - Mc \,
\sqrt{\tilde{g}_{\lambda\mu} ( q ) \dot{q}^{\lambda}
\dot{q}^{\mu}} 
 \label{AC3al}
\end{equation}
and with the torsion tensor
$\tilde{S}_{\mu\nu} {}^{\lambda} ( q )$, where the
new action principle leads to a modified Euler-Lagrange equation
like (\ref{EL3}):
\begin{equation}
\label{EL5}
\frac{d}{d \tau } \, \frac{\partial \tilde{L}}{\partial
\dot{q}^{\lambda} ( \tau  )}
-\frac{\partial \tilde{L}}{\partial q^{\lambda} ( \tau  )}
=- 2  \tilde{S}_{\lambda\mu} {}^{\nu} \left( q ( \tau  ) \right)
\dot{q}^{\mu} ( \tau  ) \frac{\partial \tilde{L}}{\partial
\dot{q}^{\nu} ( \tau  )} \, .
\end{equation}
Inserting (\ref{AC3al}) in (\ref{EL5}) and defining the proper
time $\tilde{s}$ in analogy to (\ref{PT2})
\begin{equation}
\label{PPT2}
d \tilde{s} ( \tau  ) =
\sqrt{\tilde{g}_{\lambda\mu} ( q ( \tau  ) ) d q^{\lambda} ( \tau  )
d q^{\mu} ( \tau  )} \, ,
\end{equation}
we obtain
the autoparallel equation
associated with the affine connection
${\tilde{\Gamma}}_{\mu\nu}^{\,\,\,\,\,\,\lambda}
( q )$
\begin{equation}
\label{AEN}
\ddot{q}^{\lambda} ( \tilde{s}  ) +
{\tilde{\Gamma}}_{\mu\nu}^{\,\,\,\,\,\,\lambda}
( q ( \tilde{s}  ) ) \dot{q}^{\mu}
( \tilde{s} ) \dot{q}^{\nu} ( \tilde{s}  ) = 0 \, .
\end{equation}
After reexpressing this equation in terms of the original affine
connection $\Gamma_{\mu\nu}{}^{\lambda} ( q )$
with (\ref{AF2}) and defining the proper time  $s$
via the relation
\begin{equation}
\label{PT3}
\frac{d s}{d \tilde{s}} = e^{- \sigma ( q )}
\end{equation}
due to (\ref{PT2}), (\ref{M}) and (\ref{PPT2}),
we finally get the autoparallel equation (\ref{AE1}).
Let us remark that nonintegrable time
transformations of the type (\ref{PT3})
have been extensively used in solving various classical,
quantum mechanical, and stochastic problems
\cite{Kleinert4,Duru1,Duru2,Pelster1,Pelster2,Pelster3,Pelster4}.\\

Alternatively we could have also derived this result
by rewriting the Lagrangian (\ref{AC3al}) according to (\ref{M}):
\begin{equation}
\tilde{L} ( q^{\lambda}, \dot{q}^{\lambda} ) = - Mc \,
 e^{ \sigma(q)}\sqrt{g_{\lambda\mu} ( q ) \dot{q}^{\lambda}
\dot{q}^{\mu} }.
 \label{AC3'}\end{equation}
Inserting (\ref{S}) and (\ref{AC3'}) in (\ref{EL5}), all terms involving
the scalar function $\sigma ( q )$ cancel each other, thus leading
again to the autoparallel equation (\ref{AE1}).\\

An important insight is gained by considering the
special case of
a Riemann-Cartan spacetime where the entire torsion tensor
arises from
the gradient of a scalar field $\sigma ( q )$:
\begin{equation}
\label{S2}
S_{\mu\nu}{}^{\lambda} ( q ) =
 \frac{1}{2}
\left[ \delta_{\mu}^{\,\,\,\,\lambda} \partial_{\nu} \sigma ( q ) -
\delta_{\nu}^{\,\,\,\,\lambda} \partial_{\mu} \sigma ( q ) \right] \, .
\end{equation}
>From the new action principle then follows the autoparallel equation
\begin{eqnarray}
\label{TT}
\ddot{q}^{\lambda} ( s ) \!+\!
\overline{\Gamma} {}_{\mu\nu}^{\,\,\,\,\,\,\lambda}
( q ( s ) ) \dot{q}^{\mu} ( s )
\dot{q}^{\nu} ( s ) = - \dot \sigma ( q ( s ) )  \dot{q}^{\lambda} ( s )
\!+ \!g^{\lambda\kappa} ( q ( s ) ) \partial_{\kappa}
\sigma ( q ( s ) ) \hspace*{0.7cm} \, .
\end{eqnarray}
However, after performing
the geometric gauge transformation to a purely Riemannian spacetime,
the torsion scalar $\sigma ( q )$
appears as a nongeometric external field in 
the modified Lagrangian (\ref{AC3'}), so that we find precisely the same
equation of motion via
the usual action principle.

\section{Conclusion and Outlook}

We showed that the local geometry transformations
(\ref{M}), (\ref{S}) with the property (\ref{AF2}) induce mappings
between the
autoparallel trajectories in
different Riemann-Cartan spacetimes.
This represents an unusual type of symmetry
since the equations of motion remain invariant
whereas the respective actions change.
The different terms in the actions are compensated by corresponding
contributions from the endpoints of the paths
which take into account the closure failure in the presence of torsion.
We consider this novel gauge
invariance as an interesting
property
of autoparallel trajectories
which should help giving us
hints on how to construct the proper field equations
of a theory of gravitation with torsion.

\section{Acknowledgement}

We are grateful to Drs. G. Barnich, H. von Borzeskowski, A. Scha\-kel,
S.V. Shabanov and to the graduate student C. Maulbetsch for many
stimulating discussions.

\end{document}